\documentclass[conference,11pt]{IEEEtran}

\usepackage{graphicx}
\usepackage{fancyhdr}
\usepackage{pst-node,pst-text,pst-3d,pst-char}

\usepackage{siunitx}
\usepackage{multirow}
\usepackage{dblfloatfix}
\usepackage{cite}
\usepackage{array, booktabs}
\usepackage{tikz}
\usepackage{pgfplots}
\newcolumntype{P}[1]{>{\centering\arraybackslash}p{#1}}
\newcolumntype{M}[1]{>{\centering\arraybackslash}m{#1}}

\usetikzlibrary{external}
\usepackage{textcomp}
\usepackage{lipsum}

\newcommand\copyrighttext{%
	\footnotesize \textcopyright 2020 IEEE. Personal use of this material is permitted. Permission from IEEE must be obtained for all other uses, in any current or future media, including reprinting/republishing this material for advertising or promotional purposes, creating new collective works, for resale or redistribution to servers or lists, or reuse of any copyrighted component of this work in other works.
}

\newcommand\copyrightnotice{%
    \tikzset{external/export=false}
	\begin{tikzpicture}[remember picture,overlay]
		\node[anchor=south,yshift=-3pt, xshift=10pt] at (current page.south) {\fbox{\parbox{\dimexpr\textwidth-\fboxsep-\fboxrule\relax}{\copyrighttext}}};
	\end{tikzpicture}%
	\tikzset{external/export=true}
}

\begin{document}

\title{{\large \vspace*{-3mm} \hspace*{-3mm} 2020 Fifteenth International Conference on Ecological Vehicles and Renewable Energies (EVER)}\vspace{5mm}\\
Benchmarking of a software stack for autonomous racing against a professional human race driver \vspace*{-7mm}}


\author{\IEEEauthorblockN{Leonhard Hermansdorfer, Johannes Betz, Markus Lienkamp}
\IEEEauthorblockA{
	Chair of Automotive Technology\\
	Department of Mechanical Engineering\\
	Technical University of Munich\\
	85748 Garching, Germany\\
Email: \{hermansdorfer, betz, lienkamp\}@ftm.mw.tum.de}}

\maketitle
\copyrightnotice

\begin{abstract}
The way to full autonomy of public road vehicles requires the step-by-step replacement of the human driver, with the ultimate goal of replacing the driver completely. Eventually, the driving software has to be able to handle all situations that occur on its own, even emergency situations. These particular situations require extreme combined braking and steering actions at the limits of handling to avoid an accident or to diminish its consequences. An average human driver is not trained to handle such extreme and rarely occurring situations and therefore often fails to do so. However, professional race drivers are trained to drive a vehicle utilizing the maximum amount of possible tire forces. These abilities are of high interest for the development of autonomous driving software. Here, we compare a professional race driver and our software stack developed for autonomous racing with data analysis techniques established in motorsports. The goal of this research is to derive indications for further improvement of the performance of our software and to identify areas where it still fails to meet the performance level of the human race driver. Our results are used to extend our software's capabilities and also to incorporate our findings into the research and development of public road autonomous vehicles.
\end{abstract}


\IEEEpeerreviewmaketitle

\section{Introduction}

The development of autonomous vehicles strives to increase traffic safety. By taking the human driver out of the loop, the main factor for traffic accidents is gone. Nevertheless, emergency situations will still arise, especially in mixed traffic situations. These particular situations require an extreme control input, for instance an emergency stop or sudden steering-action, in order to avoid an accident or to diminish its consequences. It is essential for an automated driving software to utilize the maximum possible tire forces and to control the vehicle at its handling limits. In a non-automated vehicle, a human driver has to handle such emergency situations by himself. Since these situations rarely happen in public road traffic, an average human driver is not trained to drive a vehicle at its handling limits. He typically fails to utilize the full tire force potential and to keep the vehicle under control. In comparison, a race driver is trained for handling such conditions. Race drivers are able to keep the vehicle under control, even in situations of slightly overshooting the handling limits. These capabilities go beyond what can be expected of an average human driver. Race driver analysis might give valuable indications for development of autonomous driving software capable of handling such extreme maneuvers in a safe way.
\newline
Our team at the Technical University of Munich develops software for a full-scale electric autonomous race car in order to participate in the Roborace Championship. \cite{Betz.2019, Betz2019b, Betz.2019c} provide an overview of our overall software concept, the idea behind Roborace and the autonomous racing vehicle itself. Parts of the software stack are available as open-source code on Github~\cite{TUM_OS.2019}.

Within the scope of this paper, we outline the most important key performance indicators~(KPI) of fast and competitive driving and how our software compares to a professional human race driver. This is done by analyzing data collected on an electric autonomous race car, which can be driven by a human and by software. This research aims to derive indications for further improvement of the performance of our autonomous driving software and to identify areas where our software still fails to meet the performance level of the human race driver.

\section{State of the art}

This section gives an overview of methods for driving style analysis in general and highlights methods with the focus on race driver evaluation. Furthermore, the latest research in terms of autonomous racing is outlined and our recent work is set into context.

\begin{table*}[b!]
	\centering{
	\caption{Five categories of driver evaluation for each driver action proposed by~\cite{Segers.2014, Trzesniowski.2017d}.}
 	\label{table:drivereval_categories}
 	
 	\renewcommand{\arraystretch}{1.5}
 	
		\begin{tabular}{|M{1.5cm}|M{2.7cm}|M{2.7cm}|M{2.7cm}|M{2.7cm}|M{2.7cm}|}
			
			\hline & Performance & Smoothness & Response & Efficiency & Consistency\\
			
			\hline Acceleration 
			& average throttle position histogram
			& throttle speed
			& full throttle point at corner exit; coasting in between; coming of the brakes and applying throttle
			& ratio positive longitudinal acceleration to throttle
			& \multirow{9}{2.7cm}{\centering evaluation of performance, smoothness and response for different corners, laps, or tracks} 
			\\
			
			\cline{1-5} Braking 
			& maximum total brake pressure; minimum longitudinal acceleration; braking point location; braking length
			& brake release smoothness
			& braking aggression; coasting between off throttle and on brakes
 			& ratio negative longitudinal acceleration to brake pressure
			& 
			\\
			
			\cline{1-5} Gear Shifting 
			& shift point; upshift duration
			& throttle blipping on downshifts
			& 
			&
			& 
			\\ 
			
			\cline{1-5} Steering 
			& driving line against lap time variance
			& steering smoothness
			& steering speed
	 		& 
			& 
			\\
			
			\hline
			
	\end{tabular}}

\end{table*}

Analysis of the driving style in public road traffic is conducted for many different purposes. \cite{Meiring.2015}~has conducted a review on driving style analysis techniques based on machine learning and artificial intelligence and has listed some of these applications. For instance, there are algorithms which aim to guide the driver towards fuel-efficient driving by evaluating the driving style with the focus on energy efficiency. This is a feature nearly all modern vehicles provide via their onboard computer. Furthermore, there are algorithms which evaluate the driving style concerning driver distraction detection and driver drowsiness detection to increase traffic safety. Another application is the driving style rating of insurance companies to adjust the insurance fee in real-time~\cite{Meiring.2015}.
\newline
Driving style evaluation is often performed with classification algorithms. \cite{Meiring.2015}~has introduced four driving style categories: normal/safe driving, aggressive driving, inattentive, and drunken driving. An important aspect in driving style classification is to not only examine the ego-vehicle but also its interaction with surrounding traffic participants, e.g., actions like tailgating~\cite{Sagberg.2015}.

With growing automation of the driving task, driving style analysis becomes increasingly obsolete, because a human is no longer in the loop. However, analyzing the driving style of the software remains an important aspect during the development process. The evaluation focuses on topics concerning passenger comfort~\cite{Bellem.2018} and on how the driving style affects surrounding traffic~\cite{Meiring.2015}.
With lower levels of automation, there are situations where the driver still has to take over the driving task. In these situations, analyzing the driver when taking over the control after a long period of disengagement is important for safety. For instance, these analyses give indications on how to adjust the time necessary for the driver to take over safely~\cite{Radlmayr.2014},~\cite{Zeeb.2015}.

The evaluation of race drivers is very different compared to evaluating a driver or software in public road traffic situations. Whereas the latter takes a variety of different drivers into account (e.g., experienced and unexperienced drivers), considers driver distraction and drowsiness, divides into safe and aggressive driving, and assesses fuel-efficient driving, the former has a limited range of interest focusing on a few key categories. The main goal of race driver analysis is to maximize vehicle and driver performance and to ultimately reduce lap time. The methods to evaluate the combination of vehicle and race driver range from analyzing the temporal course of single sensor signals to calculating a variety of KPIs. An overview of driver evaluation categories with examples is given in Table~\ref{table:drivereval_categories}, where the driver action is split into \textit{Acceleration}, \textit{Braking}, \textit{Gear Shifting}, and \textit{Steering}~\cite{Segers.2014}. In addition to the driver evaluation categories \textit{Performance}, \textit{Smoothness}, \textit{Response}, and \textit{Consistency}, we added a category named \textit{Efficiency} as introduced in~\cite{Trzesniowski.2017d}.
\newline
The approach of analyzing the temporal course of raw sensor signals or calculated signals (math channels) focuses on a specific driver action at any time~\cite{BobKnox.2011, Fey.1993}. This allows the examination and evaluation of every action and reaction a driver applies; for instance, opposite lock (also ``counter-steering") is applied via the steering wheel to react to oversteering behavior of the vehicle. Another important aspect of this analysis is to locate specific driver actions, e.g., where a driver starts to brake before entering a corner~\cite{Worle.2018}. This approach provides detailed insights into driver behavior and provides information on how to improve driver performance~\cite{BobKnox.2011, Fey.1993}. Calculated KPIs are used to assess a corner, a whole lap, or multiple laps~\cite{Trzesniowski.2017d}. This allows a comparison among laps a single driver has completed or among different drivers.

The evaluation of an autonomous vehicle driving around a race track was already examined at Stanford University~\cite{StanfordEngineering.2019}. They used a public road vehicle to drive at the limits of handling and they compared their software to a human race driver. They have concluded that the human race driver constantly exceeds and thus tests the limits of friction whereas the automated vehicle tries to follow a precalculated trajectory as close as possible~\cite{Kegelman.2018}. The control software wants to strictly follow the given trajectory and every deviation from this path is seen as control error. However, a human driver seems to purposely accept a certain range of deviation off the optimal racing line in favor of permanently probing the vehicle limits~\cite{Kegelman.2018}. By comparing open-source data from different expert race drivers, \cite{Kegelman.2017}~states that they show a high repetition accuracy at certain path points of the race track on the one hand, but significantly different distances traveled and speeds attained on the other hand. These differences are a result of different driving styles. Nevertheless, these differences in driven trajectories barely affect section and overall lap time~\cite{Kegelman.2017}.

\section{Methodology}

The analysis of race driver/ software performance is split into two sections. The first section focuses on the individual assessment of the respective ``driver". This includes highlighting individual features, especially when the software is driving the vehicle. The second section focuses on relevant KPIs that are necessary to benchmark our software against the human race driver. Individual assessment together with the KPIs is used to derive indications for further improvement of our software and to identify areas for future work. Because our race car has an electric powertrain with fixed transmission ratio, we neglect the evaluation of \textit{Gear Shifting} as one of the driver actions specified in Table~\ref{table:drivereval_categories}.

\subsection{Individual Assessment of the ``Drivers"} 
The analysis techniques for assessing a human race driver are already outlined and well established. The evaluation of a software driving a race car on a circuit autonomously is based mainly on analysis techniques already established but also has some specific features. For individual assessment with focus on specific driver actions at any time, the temporal course of raw sensor signals or calculated math channels is mainly analyzed.
\newline
A major difference to the human driver is that we know exactly what the software plans and is supposed to do. 
Hence, in our analysis we are able to constantly compare the planed trajectory to the actual one. This comparison can not be done with a manually driven vehicle, since the exact plan (i.e., planed trajectory) of the human race driver is not known. Therefore, it is possible to assess and evaluate the automated vehicle in more detail. Aside from this, many factors exist which influence the performance of an autonomous vehicle, but not of a human driven one. An example is localization, which is vital for autonomous driving since the vehicle needs to know precisely where it is located. The requirements for accuracy of the localization are even higher for a racing application. If the localization accuracy is low, the vehicle has to stick to larger safety margins or reduce the acceleration limits which directly affects lap time. These characteristics are not relevant for human driving, but have to be considered when evaluating the software’s performance. These additional factors are not a measure of performance, but have to be checked for their quality because of their impact on performance.

Furthermore, an autonomous driving software could show features in its behavior which are not covered by the established methods in a racing context. Therefore, we define additional metrics which are relevant for evaluating the autonomous driving software performance:

\begin{itemize}
	\item deviation of planned trajectory: target trajectory is defined as vehicle position, heading and velocity. The difference between actual and target trajectory is used for evaluation of the software's tracking accuracy.
	\item oscillations originating from within the plan/control/act pipeline: this category aims to identify oscillations induced within the pipeline from trajectory planning to execution via the respective actuators, i.e., throttle, brake and steering. The planed trajectory provides target values which are forwarded to the control software. The controller calculates a longitudinal force request which is then translated into engine torque or brake pressure and a target curvature which is requested as steering angle. Afterwards, the respective actuators have to enforce these requested values. We compare the smoothness of the handed-over values between planning and control, as well as the actuators in order to identify sources of oscillations which would harm vehicle performance.
	\item sensor signal quality: the software functionality relies heavily on the quality of sensor signals, e.g., for localization. Therefore, decreasing sensor signal quality or a sensor failure have to be considered as a cause for performance loss. Indications for decreasing sensor quality are, for instance, an increased noise level or drifting rate of signals.
\end{itemize}

These signals can be averaged over individual laps to use them as \textit{autonomous racing KPIs}. This allows to compare multiple laps and runs of the software. However, a comparison with a human driver is not reasonable.

\subsection{KPIs for Performance Assessment} 
In the following, a selection of different KPIs for evaluation of the driver actions throttle, braking, and steering are outlined. In addition, KPIs to evaluate vehicle stability and driving line are presented.

\subsubsection{throttle} we calculate an artificial throttle signal for the autonomous vehicle, based on the longitudinal force demand of the controller, because the software does not provide a throttle signal as it is available with a human driver. During human driving, the throttle is applied via the throttle pedal.

\begin{itemize}
	\item throttle acceptance: proportion of the lateral acceleration at the exit of a corner, at the time when the driver applies full throttle, to the maximum lateral acceleration of the respective corner. It is a measure of traction and driving style, the value should be as high as possible~\cite{Segers.2014}.
	\item coasting time: total time where neither brakes nor throttle is applied. In general, this value should be kept as low as possible to achieve a fast lap time~\cite{Segers.2014}.
\end{itemize}

\subsubsection{braking} both brake pressure signals (front and rear axle) are used for evaluation of the braking action. 
\begin{itemize}
	\item brake pressure aggression: time derivative of the sum of front and rear brake pressure signals as a measure of how fast the driver applies the brakes (only positive time derivative). A high value indicates good braking technique~\cite{Segers.2014, Trzesniowski.2017d}.
	\item brake release smoothness: a measure of how smooth the driver comes off the brakes. Releasing the brakes is not only about quickness but also about adequately modulating the brake pressure to account for changing vehicle behavior under braking, e.g., decreasing downforce~\cite{Segers.2014}. This value is calculated the same way as brake pressure aggression, but only negative values are taken into account, which corresponds to the action of releasing the brakes and decreasing brake pressure. A high value means that the brake pressure is reduced too quickly~\cite{Trzesniowski.2017d}.
	\item braking quickness: time between start of braking action and first negative longitudinal acceleration peak; should be as low as possible~\cite{Segers.2014}.
\end{itemize}

\subsubsection{steering} the steering angle, which is not the steering wheel angle but the averaged angle of both front wheels, is used for the evaluation of the steering action.
\begin{itemize}
	
	\item steering speed/smoothness: steering speed is the time derivative of the steering angle signal $\delta_{steer}$. Steering smoothness $K_{SS}$ is the difference between the measured steering angle and the smoothed steering signal~\cite{Trzesniowski.2017d}. It provides indications on how aggressive the driver is steering, and also if abrupt steering corrections are necessary due to oversteering vehicle behavior~\cite{Segers.2014}.
 	\begin{equation}
	K_{SS} = \left|\delta_{steer,raw} - \delta_{steer,smoothed}\right| \label{eq:steering_smoothness}
 	\end{equation}
	\item steering integral: the absolute value strongly depends on the race track, as longer tracks result in a higher value. Differences between laps of the same track arise from steering corrections and choice of the driven path. A reference lap, recorded to get an ideal steering angle, provides a reference value for the steering angle integral. Based on this reference, a higher value of steering angle integral hints at an understeering vehicle, where the driver has to apply additional steering action. If the vehicle tends to oversteer, the driver has to steer in the opposite direction resulting in a smaller value compared to the reference lap~\cite{Trzesniowski.2017d}.
 	\begin{equation}
	K_{SI} = \int_{t_0}^{t_{lap}} \left|\delta_{steer}\right| dt \label{eq:steering_integral}
 	\end{equation}
	
\end{itemize}

\subsubsection{vehicle stability} the focus of vehicle stability analysis is on lateral vehicle behavior, mainly on understeering and oversteering.
\begin{itemize}
	
	\item attitude velocity: difference between measured yaw rate $\dot{\psi}_{measure}$ and angular velocity $\omega_{ang}$, with lateral acceleration $a_y$ and longitudinal velocity $v_x$. Positive values suggest oversteering and negative values understeering behavior~\cite{Segers.2014}.
 	\begin{equation}
	\Delta \dot{\psi} = \dot{\psi}_{measure} - a_y/v_x \label{eq:delta_dpsi}
 	\end{equation}
	\item delta tire slip angle: difference between the averaged tire slip angle at the front and rear axle. Positive values suggest understeering and negative values oversteering behavior~\cite{Trzesniowski.2017d}.
 	\begin{equation}
 		\Delta \alpha = \alpha_{front} - \alpha_{rear} \label{eq:delta_aplha}
 	\end{equation}
 	
\end{itemize}

\subsubsection{driving line} lateral deviation of the race trajectory which, at the same time, is the target trajectory of the autonomous vehicle. As previously mentioned, this trajectory can not be used to evaluate the race driver in terms of path matching. By using it as a reference trajectory, it is useful for comparing both drivers' driven trajectories.

\section{Results}

\begin{figure}[thpb]
	\centering
	
 	\resizebox{0.495\textwidth}{!}{
	\input{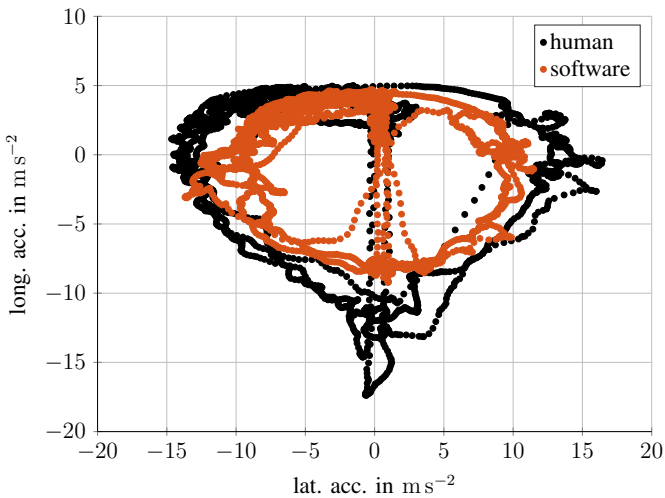}
	}
	\caption{Overlay of the vehicle acceleration data of a human race driver (black) and our software (orange).}
	\label{fig:results_tractioncircle}
	
\end{figure}

In this section, we analyze two laps around a racetrack (Monteblanco Circuit in Spain), recorded in March 2019. One lap was driven by a human race driver and the other one by our software. Each lap was the fastest of the respective ``driver". The human driver was able to familiarize with the track beforehand, and we conducted a mapping run to obtain the driveable space for trajectory planning. Each run consisted of multiple laps for tire warm-up. The comparison is done by analyzing the temporal course of sensor signals and math channels. This allows the evaluation of the differences in driver actions at every time step, and for specific situations like corner entry and corner exit. In addition, lap-based KPIs give indications on how both drivers perform over one complete lap.
Figure~\ref{fig:results_tractioncircle} shows the overlay of both g-g diagrams, i.e., longitudinal and lateral acceleration of the whole lap.

\begin{figure}[thpb]
	\centering
	
 	\resizebox{0.495\textwidth}{!}{
		\input{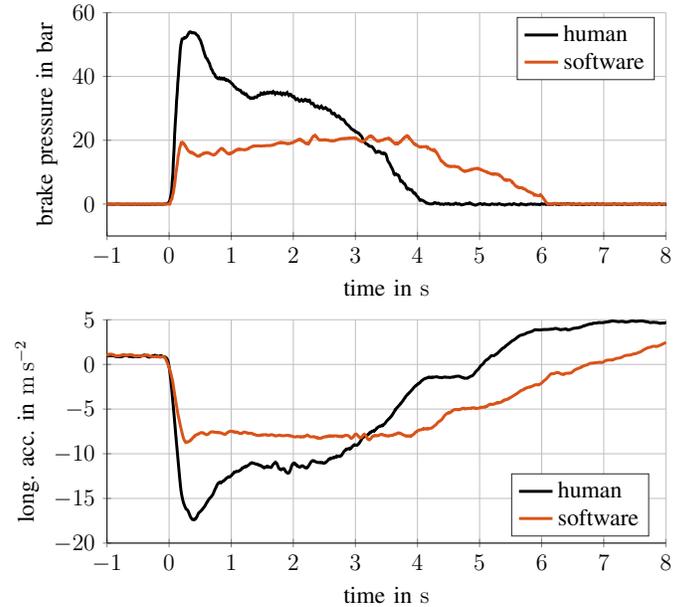}
	}
	\caption{Overlay of the drivers' brake pressure application and the resulting longitudinal vehicle acceleration; data of human race driver (black) and our software (orange).}
	\label{fig:results_brakecomp}
	
\end{figure}

The overlay of both g-g diagrams reveals the differences in what we extract out of the vehicle and what the driver is able to extract. We assumed the g-g diagram to be of circular shape, which also shows in the data. However, when plotting the measured acceleration, the actual g-g diagram of the race driver in negative longitudinal acceleration appears to be straight, not curved, having a diamond-like shape. We lose performance in both directions of pure acceleration but reach the vehicle dynamics limits partly in combined acceleration. At the time of data recording, aerodynamic downforce effects were not implemented in our trajectory planning, which additionally increases the gap to the human driver. This can be seen under braking. Figure~\ref{fig:results_brakecomp} shows the brake pressure and the resulting negative longitudinal acceleration of the human driver and our software. Here, the driver is able to modulate the acceleration depending on the current vehicle velocity, whereas the software sticks strictly to the predefined limit. The human driver is applying nearly three times the brake pressure leading to a significantly higher deceleration. More importantly, he lowers the brake pressure with decreasing velocity to account for velocity-dependent downforce and to avoid wheel lock-up. The software version currently does not consider this velocity-dependent downforce and applies brake pressure according to the predefined limitations. This leads to a nearly constant brake pressure and deceleration. Whereas the driver is stepping off the brakes after~\SI{4}{\second}, the automated vehicle finishes the braking action about~\SI{2}{\second} later due to conservative braking.

The throttle KPIs in Table~\ref{table:drivercomp_results} show a gap in \textit{throttle acceptance}, meaning that the driver is applying full throttle earlier, after the lateral acceleration peak when cornering, than the software. This leads to a higher combined acceleration and therefore higher utilization of the tire force potential at corner exit. The software is applying throttle more conservatively. This KPI does not consider the absolute lateral acceleration of the respective driver. It should be noted that the human driver is reaching an average peak lateral acceleration of \SI{13.8}{\meter\per\square\second}, whereas the software only reaches~\SI{10.9}{\meter\per\square\second}. The conservative throttle application results in a visible dent at the top left of the software's g-g diagram in Figure~\ref{fig:results_tractioncircle}. The human driver achieves a convex g-g diagram in this particular area. The \textit{coasting time} of the software is about~\SI{1.6}{\second} higher compared to the human driver. For both drivers, coasting happens after coming off the brakes and before applying throttle. Coasting before braking is absent.

\begin{table}[t!]
	\centering{
		\caption{Selected KPIs: human race driver and our software at the Monteblanco racetrack in Spain.}
		\label{table:drivercomp_results}
		
		\renewcommand{\arraystretch}{1.5}
		\begin{tabular}{|M{2.0cm}|M{1.6cm}|M{1.6cm}|M{1.6cm}|}
			
			\hline KPIs (lap average) & human race driver & TUM Roborace software & relative difference to human\\
			
			\hline max. velocity in \SI{}{\meter\per\second}
			& 61.5
			& 58.0
			& -\SI{6}{\percent}\\ 
		
			\hline max./min. long. acceleration in \SI{}{\meter\per\square\second}
			& 5.0/-17.4
			& 4.7/-9.2
			& -\SI{6}{\percent}/ -\SI{47}{\percent}\\ 
			
			\hline max. lat. acceleration in \SI{}{\meter\per\square\second}
			& 16.5
			& 13.7
			& -\SI{17}{\percent}\\ 
			
			\hline lap time in \SI{}{\second}
			& 63.03
			& 69.98
			& +\SI{11}{\percent}\\ 
			
			\hline throttle acceptance in \SI{}{\percent}
			& 75
			& 33
			& -\SI{56}{\percent}\\
			
			\hline coasting time in \SI{}{\second}
			& 1.99
			& 3.65
			& +\SI{83}{\percent}\\ 
			
			\hline brake pressure aggression in \SI{}{bar\per\second}
			& 371
			& 259
			& -\SI{30}{\percent}\\
			
			\hline brake release smoothness in \SI{}{bar\per\second}
			& 38.9
			& 10.8
			& -\SI{72}{\percent}\\
			
			\hline braking quickness in \SI{}{\second}
			& 0.28
			& 0.25
			& -\SI{10}{\percent}\\
			
			\hline steering speed in \SI{}{\radian\per\second}
			& \SI{1.39e-4}
			& \SI{6.94e-5}
			& -\SI{50}{\percent}\\
			
			\hline steering integral in \SI{}{\radian\second}
			& 1.93
			& 1.74
			& -\SI{10}{\percent}\\
			
			\hline attitude velocity in \SI{}{\radian\per\second} $(\Delta \dot{\psi} >0/<0)$
			& 0.045/ -0.039
			& 0.030/ -0.027
			& -\SI{33}{\percent}/ -\SI{31}{\percent} \\			
			
			\hline delta tire slip angle in \SI{}{{\radian}} $(\Delta \alpha >0/<0)$
			& 0.014/ -0.007
			& 0.004/ -0.005
			& -\SI{71}{\percent}/ -\SI{29}{\percent}\\
			
			\hline lateral deviation of the race trajectory in \SI{}{\meter}
			& 0.88
			& 0.30
			& -\SI{66}{\percent}\\
			
			\hline
			
	\end{tabular}}
	
\end{table}

The brake KPIs in Table~\ref{table:drivercomp_results} show that the human driver is applying the brakes quicker~(\textit{brake pressure aggression}), which is also visible in the steeper slope of the brake pressure in Figure~\ref{fig:results_brakecomp}. The human driver does release the brakes quicker than the software~(\textit{brake release smoothness}). In general, a higher value of this KPI is seen as bad braking technique\cite{Trzesniowski.2017d, Segers.2014}. However, since the software shows a significantly different, conservative braking technique, a valid comparison is not possible. This is also true for \textit{braking quickness}, because the software reaches lower maximum deceleration than the human driver. Both KPIs become relevant when both performance levels converge in the future.

\begin{figure}[bp!]
	\centering
	
	\resizebox{0.495\textwidth}{!}{
		\includegraphics{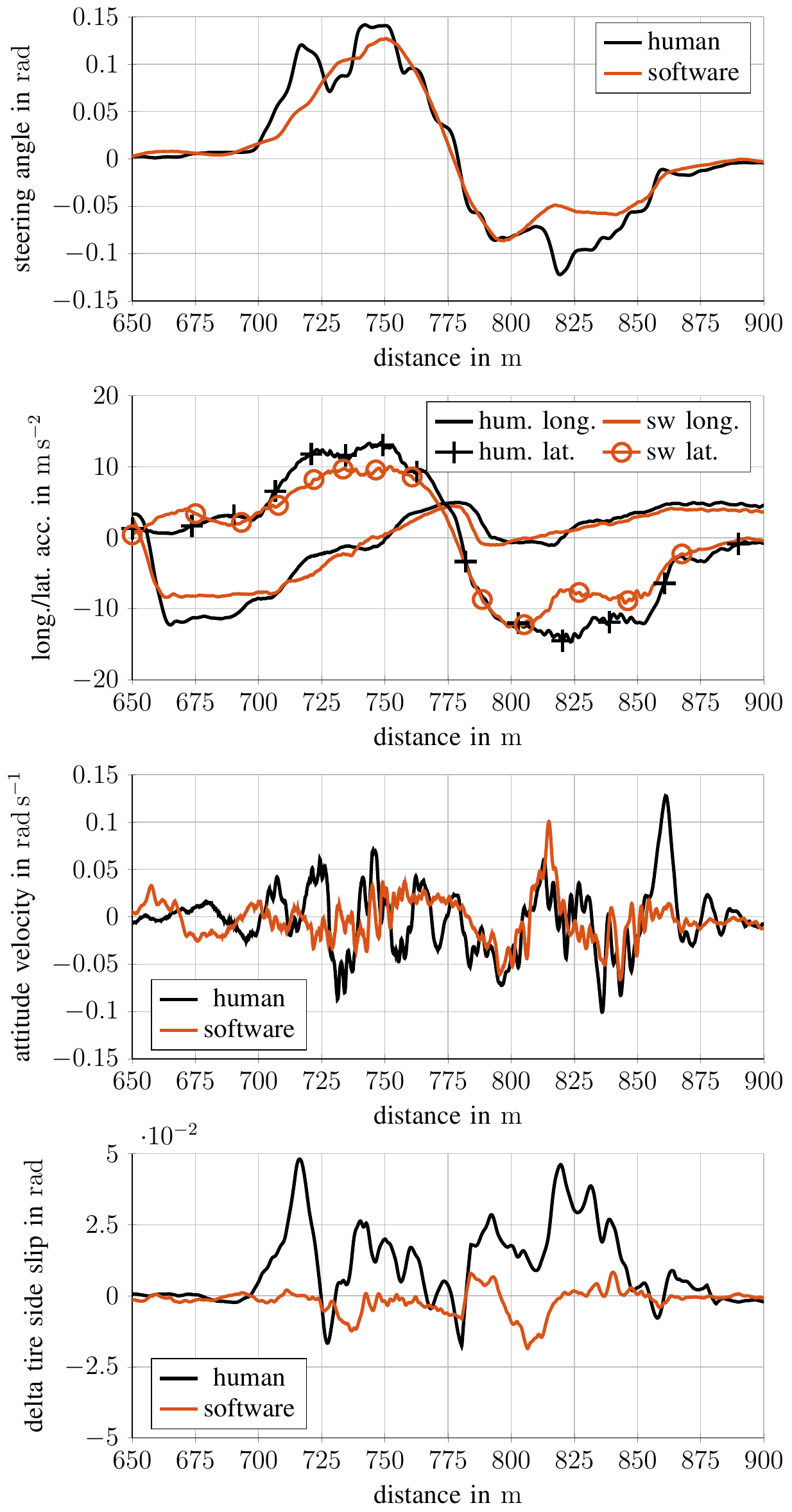}
	}
	\caption{Overlay of steering angle, long./lat. vehicle acceleration, attitude velocity and delta tire slip angle; data of human race driver (black) and our software (orange).}
	\label{fig:results_steercomp}
	
\end{figure}

The steering KPIs in Table~\ref{table:drivercomp_results} provide information on lateral vehicle stability. \textit{Steering speed} is the absolute value of the steering angle time derivative. Next to influencing factors such as velocity or driving line, higher average values indicate an increased amount of steering corrections the driver has to apply~\cite{Trzesniowski.2017d}. Figure~\ref{fig:results_steercomp} shows the steering angle of both drivers during a corner. The software is applying a smooth steering angle throughout the corner resulting in an overall smaller \textit{steering speed}. In comparison, the human driver's steering angle is more agitated, e.g., around~\SI{725}{\meter}, \SI{825}{\meter}, and \SI{860}{\meter} where he corrects his steering angle abruptly. These abrupt steering corrections result in a higher average \textit{steering speed}. 
The absolute value of the \textit{steering integral} heavily depends on the actual track layout and provides no valuable information. However, this KPI can be useful for comparison, because it represents a measure of how much steering the driver has to apply. A higher value means that the driver has to steer more to overcome understeering whereas a lower value suggests an oversteering tendency~\cite{Trzesniowski.2017d}. Because the vehicle in automated driving mode is farther away from the handling limit than the human driver, it experiences less over- and understeering behavior. Therefore, with the \textit{steering integral} being larger for the human driver, this suggests an increased understeering behavior.

The vehicle stability KPIs in Table~\ref{table:drivercomp_results} also indicate that the human driver encounters more over- and understeering behavior than the automated vehicle. In Table~\ref{table:drivercomp_results}, average values for \textit{attitude velocity} and \textit{delta tire slip angle} are calculated for negative and positive values. A positive value provides a measure for oversteering (\textit{attitude velocity}) and understeering (\textit{delta tire slip angle}), and vice versa. In general, the values show that the human driver is experiencing more over- and understeering behavior, but the values itself do not provide a distinct observation, e.g., giving a difference in understeering behavior of \SI{-31}{\percent} and \SI{-71}{\percent}. The math channels on which these KPIs are based can provide useful information when assessing specific driver action and vehicle behavior in detail. Figure~\ref{fig:results_steercomp} shows both vehicle stability math channels over a specific track distance in comparison. A clear oversteering tendency can be seen at around~\SI{725}{\meter} and \SI{860}{\meter} (both human driver), and around~\SI{810}{\meter} (software). The overall understeering tendency in \textit{delta tire slip angle} of the human driver cannot be backed with \textit{attitude velocity}. Between \SI{700}{\meter} - \SI{725}{\meter} and \SI{810}{\meter} - \SI{830}{\meter} of the human driver especially, both values show contrary behavior. As a conclusion, both math channels over time are useful for detailed driver assessment. However, their lap-based KPIs seem to provide no reliable absolute value but only a correct tendency.

\begin{figure}[t!]
	\centering
	
	\resizebox{0.48\textwidth}{!}{
		\input{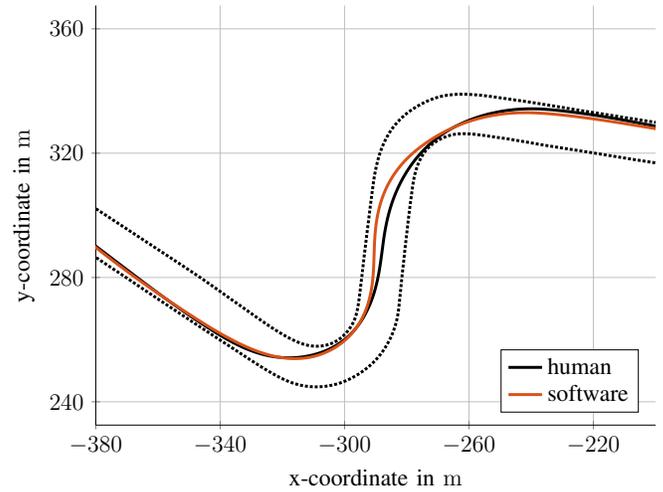}
	}
	\caption{Overlay of the driving line of the human race driver (black) and our software (orange).}
	\label{fig:results_pathcomp}
	
\end{figure}

The lateral deviation of the human driver is higher compared to the software, but does not provide valuable information on which driving line is better. Whereas an evaluation of the tracking performance of the autonomous vehicle is possible, because the reference path is the target path, we cannot rate the human driver using the lateral deviation of this reference path. Nevertheless, comparing both actual driving lines provides information on different driving styles and proves important for detailed driver analysis. Figure~\ref{fig:results_pathcomp} depicts both driving lines for the above examined track section between \SI{650}{\meter} and \SI{900}{\meter}. Both drivers enter the first corner in the same way in terms of driving line. The second corner is handled in different ways. The human driver applies less steering leading to a straighter path between both corners. At corner entry the human driver takes an early apex line, where the driver forces the vehicle to complete the corner after the actual apex~\cite{Fey.1993}. Our software's trajectory planner minimizes curvature and therefore drives a mid-apex line, which deviates significantly from the race driver's driving line at corner entry. Additionally, the human driver includes the curbs at the corner inside into his driving line. The software stays off the curbs, which is enforced by predefined safety distances in trajectory planning.

The \textit{autonomous racing KPIs} are not explicitly outlined, because they do not show any anomalies which would decrease the vehicle's performance.

\section{Conclusion And Future Work}

This research outlines the specific features of our automated driving software near the handling limits compared to a professional human race driver. The comparison shows aspects where our software is not able to reach the human race driver's level of performance and provides clear indications on which aspects we have to focus on to close this gap.

The human driver still significantly outperforms our software, the main reason being different acceleration limits. Our software uses conservative, velocity-independent acceleration limits for trajectory planning. In addition, imposed safety distances limit the vehicle's ability to compete. Our software follows the race trajectory more smoothly and requires less steering corrections which is the result of not driving at the handling limits. In the future, our software has to be able to control the vehicle at higher acceleration, which requires fast and adequate corrective actions in both the longitudinal and lateral direction. The main aspects for future work are: online adaption of the safety distances, online assessment of the vehicle performance limit considering velocity-dependent influences, and robust controlling of the vehicle at the handling limits. These topics should enable the software to narrow the gap to the race driver’s skills of constantly driving at the handling limits as well as adapting quickly to changing environmental conditions.
\newline
Furthermore, our goal is to automate the process of data analysis and interpretation, which is human-centered in its current state, and to communicate the findings directly to the autonomous driving software. Data analysis should not only be conducted offline after each run, but also on-board the vehicle.

The presented data was recorded in March 2019. Since this time, we have updated our software with the goal of reducing the gap to the human driver. One major change is altering the shape of the traction circle used for trajectory planning, as already mentioned. This allows us to use the maximum tire forces much better than before. We have already tested the upgrades in simulation and on other tracks. Data of a human race driver on these tracks is not available, however, a final evaluation of the software in comparison to a human race driver is planned.

\section*{Acknowledgments and Contributions}

Research was supported by the Bavarian Research Foundation (BFS). 
As the first author, Leonhard Hermansdorfer initiated the idea of this paper and is responsible for the presented concept and its implementation. His contribution was essential to the overall design of the data analysis software and its automation on the race car. Johannes Betz contributed to the overall concept and system design. Markus Lienkamp contributed to the conception of the research project and revised the paper critically for important intellectual content. He gave final approval of the version to be published and agrees to all aspects of the work. As guarantor, he accepts responsibility for the overall integrity of the paper.

\bibliographystyle{ieeetr}
\bibliography{EVER2020_humanVSsw_biblio}

\end{document}